\documentclass[
]{ceurart}

\usepackage{listings}
\usepackage[inline]{enumitem}

\newcommand{\lilybench}{\textsc{LilyBench}}
\newcommand{\phifour}{Phi-4}
\newcommand{\qwencoder}{Qwen2.5-Coder-14B}
\newcommand{\deepseekcoder}{DeepSeek-Coder-V2-Lite}
\newcommand{\codestral}{Codestral-22B}

\begin{document}

\copyrightyear{2026}
\copyrightclause{Copyright for this paper by its authors.
  Use permitted under Creative Commons License Attribution 4.0
  International (CC BY 4.0).}

\conference{Ital-IA 2026: 6th National Conference on Artificial Intelligence, organized by CINI, June 18-19, 2026, Rome, Italy}

\title{Can LLMs understand LilyPond? A benchmark for symbolic music generation and understanding}


\author[1]{Matteo Spanio}[orcid=0000-0002-5780-2456,email=spanio@dei.unipd.it]
\cormark[1]
\author[1]{Mohammad Torabi}[email=mohammad.torabi@studenti.unipd.it]
\author[2]{Andrea Poltronieri}[orcid=0000-0003-3848-7574,email=andrea.poltronieri@upf.edu]
\author[1]{Antonio Rod\`a}[orcid=0000-0001-9921-0590,email=roda@dei.unipd.it]

\address[1]{Centro di Sonologia Computazionale, University of Padova, Padova, Italy}
\address[2]{Music Technology Group, Universitat Pompeu Fabra, Barcelona, Spain}

\cortext[1]{Corresponding author.}

\begin{abstract}
Symbolic music evaluation for large language models remains fragmented across representations, datasets, and metrics. We introduce LilyBench, a LilyPond-based benchmark that jointly evaluates symbolic music generation and music understanding on the same family of open-weight LLMs. The benchmark includes a 200-prompt generation suite and ten understanding tasks adapted from ABC-Eval, covering syntax, metadata prediction, structural sequencing, and music recognition. Generation quality is evaluated using compile rate, MusPy descriptor distributions via Jensen–Shannon similarity, and LilyBERT-based Fréchet Music Distance (FMD). Experiments on four open-weight models show that executable LilyPond generation is achievable in zero-shot settings, while structural understanding tasks remain challenging despite strong performance on composer and genre recognition. Our experiments also reveal systematic disagreements between descriptor-based and embedding-based metrics, suggesting that symbolic music evaluation benefits from metric triangulation rather than single-score ranking. We release the benchmark, prompt bank, and evaluation code to support future research in symbolic music generation and understanding at \url{https://github.com/CSCPadova/lilybench}.
\end{abstract}


\maketitle

\section{Introduction}
The choice of music representation determines which constraints a language-model system can check automatically and which musical knowledge stays latent~\cite{ma2024foundation, spanio2025frontiers}. Among textual score formats, ABC notation has received the most attention, both for generation~\cite{wang2025notagen,casini2024investigating} and understanding~\cite{zhao2026abceval,yuan2024chatmusician}; LilyPond~\cite{lilypond2026} is less studied, despite exposing a richer score language closer to code that fits full-score generation and structural reasoning. The BMdataset corpus and LilyBERT encoder~\cite{spanio2026bmdataset} make it a practical substrate for LLM work.

We ask whether current open-weight LLMs can generate well-formed LilyPond matching a curated Baroque reference, whether the same models can answer structural, metadata, style, and quality questions from raw LilyPond, and how descriptor-based (Jensen--Shannon) and embedding-based (Fr\'echet Music Distance) distributional metrics differ when ranking generation quality. To answer these, we introduce \lilybench{}, a benchmark that pairs a generation evaluation with an understanding suite of ten tasks adapted from ABC-Eval~\cite{zhao2026abceval}, evaluated on the same four open-weight backbones: \phifour{}, \qwencoder{}, \deepseekcoder{}, and \codestral{}.

Our contributions are:
\begin{enumerate*}[label=(\arabic*)]
  \item the first symbolic music evaluation framework based on \emph{LilyPond}, enabled by BMdataset and LilyBERT~\cite{spanio2026bmdataset};
  \item the first \emph{unified evaluation} of LLM music generation \emph{and} understanding on the same backbones; and
  \item to the authors' knowledge, the first head-to-head comparison of \emph{Jensen--Shannon descriptor similarity} and \emph{Fr\'echet Music Distance} on the same reference corpus, showing that the two metrics capture complementary failure modes.
\end{enumerate*}

\section{Related work}
MIDI is the dominant symbolic representation for music generation~\cite{le2025survey}, but it is performance-oriented and exposes neither score structure nor variable reuse. Among text-based formats, ABC has received the most attention: as a ``second language'' for LLaMA-2~\cite{yuan2024chatmusician}, in full pretrain/finetune/RL pipelines~\cite{wang2025notagen}, with Russell-quadrant emotion conditioning~\cite{zhou2024emelodygen}, and as the substrate for the ABC-Eval understanding benchmark~\cite{zhao2026abceval}. LilyPond~\cite{lilypond2026} has typically been used as a rendering target rather than a model substrate; the BMdataset corpus and LilyBERT encoder~\cite{spanio2026bmdataset} make it usable for LLM work. Objective evaluation is more mature for raw audio, where Fr\'echet Audio Distance~\cite{kilgour19_interspeech} and CLAP-based similarities are standard; symbolic music has lagged, mixing parseability checks, handcrafted descriptors, and manual assessment~\cite{kader2025survey}. Current symbolic evaluation falls into two distributional families. Descriptor distributions aggregate MusPy's twelve symbolic features~\cite{dong2020muspy} via Jensen--Shannon divergence~\cite{qian2026pianoroll}. Embedding distributions use symbolic encoders such as MusicBERT~\cite{zeng2021musicbert} and CLaMP~\cite{wu2023clamp} as spaces in which Fr\'echet Music Distance (FMD)~\cite{retkowski2024fmd} ports the FID/FAD construction. To our knowledge, no prior work has run JS and FMD head-to-head on the same models and corpus. Music benchmarks for LLMs consistently report a recognition-vs-reasoning gap~\cite{yuan2024chatmusician,zhou2024canllmsreason}, and the closest precedent for our understanding suite is ABC-Eval~\cite{zhao2026abceval}, a 10-task benchmark over ABC notation. \lilybench{} mirrors its task taxonomy, ports each task to LilyPond, and pairs it with a generation benchmark, so the same backbone is evaluated on both sides from the same notation.

\section{Methodology}
\lilybench{} couples a generation benchmark with an understanding benchmark, both run on the same four backbones and reference corpus.

We evaluate four open-weight backbones, one general-purpose and three code-specialised. \phifour{}~\cite{abdin2024phi4} (14~B dense) is the general-purpose baseline. The code-centric checkpoints are \qwencoder{}~\cite{hui2024qwencoder} (14~B), \deepseekcoder{}~\cite{zhu2024deepseekcoderv2} (16~B MoE, $\approx$2.4~B active), and \codestral{}~\cite{mistral2024codestral} (22~B). Sizes are not matched; we test whether code pretraining transfers to LilyPond.

The reference corpus is BMdataset~\cite{spanio2026bmdataset}: 391 Baroque works yielding 2{,}645 self-contained LilyPond files. We split at the work level to prevent leakage; the held-out test split is the in-domain reference and Mutopia the out-of-domain one. Understanding additionally uses EMOPIA for emotion (120 records over Russell quadrants) and synthetically corrupted Mutopia scores for error detection (220 records over five corruption categories).

\subsection{Generation setup}
All cells use the same 200-prompt stratified bank. \emph{Zero-shot}: a short natural-language instruction plus a metadata block (composer, period, form, ensemble, part). \emph{Few-shot}: three demonstrations sampled from the training distribution. \emph{Few-shot (Ablation)}: three hand-written A-minor examples, retained because they expose a JS/FMD disagreement (section \ref{sec:results}).

\subsection{Understanding tasks}
We adapt the ten ABC-Eval tasks~\cite{zhao2026abceval} to LilyPond, grouped by the original three-level reasoning depth (Table~\ref{tab:understanding-tasks}). Titles are extracted from \texttt{\textbackslash header}; the \texttt{L:1/N} note length is read from the \texttt{\textbackslash time} signature; distractors come from the same-corpus pool or within-score bars; bench JSONLs are emitted under a fixed seed. Decoding is greedy ($T=0$, \texttt{max\_new\_tokens=20}) with no chain-of-thought, following ABC-Eval.

\begin{table}[t]
\centering
\scriptsize
\setlength{\tabcolsep}{3pt}
\begin{tabular}{llccll}
\toprule
Category & Task & $n$ & Source & Format & Metric \\
\midrule
\multirow{2}{*}{Basic}
 & bar\_count           & 100 & Mutopia & integer       & exact-match acc. \\
 & metadata\_qa         &  60 & Mutopia & 4-way MC      & accuracy \\
\midrule
\multirow{3}{*}{Segment}
 & bar\_sequencing      & 119 & Mutopia & 4-digit perm. & Kendall-$\tau$ (pen.) \\
 & next\_bar\_prediction& 119 & Mutopia & 4-way MC      & accuracy \\
 & metadata\_prediction &  60 & Mutopia & 4-way MC      & accuracy \\
\midrule
\multirow{5}{*}{Sequence}
 & music\_captioning    &  60 & Mutopia        & 4-way MC          & accuracy \\
 & composer\_recognition&  96 & Mutopia        & 4-way MC          & accuracy \\
 & genre\_recognition   & 132 & Mutopia        & 4-way MC          & accuracy \\
 & emotion\_recognition & 120 & EMOPIA         & 4-way MC (Russell)& accuracy \\
 & error\_detection     & 220 & Mutopia$^\dagger$ & bar list       & macro-F1 \\
\bottomrule
\end{tabular}
\caption{The ten understanding tasks adapted from ABC-Eval~\cite{zhao2026abceval} into LilyPond, grouped by reasoning depth. Inputs are raw LilyPond text; decoding is greedy. $^\dagger$ Mutopia scores with five categories of synthetic corruption (invalid metadata/content/bar duration, melodic leap, accidental outside key).}
\label{tab:understanding-tasks}
\end{table}

\subsection{Metrics}
\label{sec:metrics}
\paragraph{Generation.} (1) \textbf{Compile rate}: text to MIDI via the LilyPond binary. (2) \textbf{JS similarity}: $100\cdot\exp(-2\bar{D}_{\text{JS}})$ over Gaussian fits to three MusPy descriptors (polyphony rate, groove consistency, scale consistency), on rendered outputs only. (3) \textbf{LilyBERT-based FMD}: $\lVert\mu_r-\mu_g\rVert^2 + \mathrm{Tr}(\Sigma_r+\Sigma_g-2\sqrt{\Sigma_r\Sigma_g})$ on layer-6 LilyBERT embeddings of raw LilyPond text~\cite{retkowski2024fmd,spanio2026bmdataset}; compile-independent. We report JS and FMD against BMdataset test (in-domain) and Mutopia (out-of-domain).

\paragraph{Understanding.} Accuracy for the six 4-way MC tasks; exact-match for \texttt{bar\_count} with $\pm 1/\pm 5/\pm 10$ tolerance breakdown; penalised Kendall-$\tau$ for \texttt{bar\_sequencing}; macro-F1 for \texttt{error\_detection}. Mutopia summaries report a macro average over the eight Mutopia tasks. Chance: 0.25 for 4-way MC; effectively 0 for \texttt{bar\_count} and \texttt{error\_detection}.

\section{Evaluation}
\subsection{Generation results}
\label{sec:results-gen}
Table~\ref{tab:matrix} reports the twelve generation cells against both references. The zero-shot baseline is already usable: \phifour{}, \qwencoder{}, and \codestral{} compile 69--79\% of prompts, and \deepseekcoder{} trails at 48.6\% but stays competitive on FMD; the leaderboard depends on the metric (\codestral{} leads on compile, \qwencoder{} on JS(test), \phifour{} on out-of-domain FMD), so no single score ranks the four models. Train-derived few-shot demonstrations pull all four closer to the reference in LilyBERT space (FMD(test) 0.696--0.742) but cost executability (compile 19.9--45.2\%), as outputs land in long-tail regions of the corpus that are syntactically more fragile. The hand-written A-minor demonstrations of the ablation regime behave differently: compile rates climb to 97--100\% but FMD(test) jumps to 1.754--1.980, and JS disagrees with FMD here, with \codestral{} reaching the table's highest JS(test) (89.44) while its FMD remains the worst in its column. We return to this disagreement in section \ref{sec:results}.

\begin{table}[t]
\centering
\scriptsize
\setlength{\tabcolsep}{3pt}
\begin{tabular}{llccccc}
\toprule
Regime & Model & Comp.~(\%) & JS(test) $\uparrow$ & JS(Mut.) $\uparrow$ & FMD(test) $\downarrow$ & FMD(Mut.) $\downarrow$ \\
\midrule
\multirow{4}{*}{Zero-shot}
 & \phifour   & 71.1 & 83.27 & \textbf{81.24} & \textbf{0.933} & \textbf{1.419} \\
 & \qwencoder & 69.0 & \textbf{84.85} & 73.80 & 1.139 & 1.681 \\
 & \deepseekcoder & 48.6 & 55.39 & 58.07 & 0.887 & 1.578 \\
 & \codestral & \textbf{79.3} & 75.78 & 65.56 & 0.960 & 1.722 \\
\midrule
\multirow{4}{*}{Few-shot}
 & \phifour  & 35.1 & \textbf{74.80} & \textbf{67.58} & 0.701 & \textbf{1.278} \\
 & \qwencoder  & 19.9 & 63.43 & 65.69 & 0.742 & 1.414 \\
 & \deepseekcoder  & 26.3 & 57.04 & 60.23 & 0.714 & 1.428 \\
 & \codestral  & \textbf{45.2} & 67.55 & 59.18 & \textbf{0.696} & 1.407 \\
\midrule
\multirow{4}{*}{Ablation}
 & \phifour  & 99.6 & 71.13 & 69.50 & 1.874 & 2.683 \\
 & \qwencoder  & 98.9 & 63.09 & 55.93 & 1.980 & 2.796 \\
 & \deepseekcoder  & 99.9 & 53.44 & 46.96 & 1.960 & 2.773 \\
 & \codestral  & 97.1 & \textbf{89.44} & \textbf{76.95} & 1.754 & 2.535 \\
\bottomrule
\end{tabular}
\caption{Generation matrix, 200 prompts per cell. \emph{Ablation} uses hand-written A-minor demonstrations. JS and FMD are computed against the BMdataset test split and the out-of-domain Mutopia reference. Bold marks the best value within each regime.}
\label{tab:matrix}
\end{table}

\subsection{Understanding results}
\label{sec:results-und}
Table~\ref{tab:understanding-results} reports per-task scores. Sequence-level recognition (\texttt{music\_captioning} 0.68--0.94, \texttt{composer\_recognition} 0.50--0.89, \texttt{genre\_recognition} 0.79--0.96) sits well above the 0.25 chance baseline, but structural syntax does not: \texttt{bar\_count} exact match is 0--3\% and \texttt{error\_detection} macro-F1 stays at 0.041 or below for every model and corruption category. \phifour{} and \qwencoder{} tie at the top (Mutopia macro 0.651, 0.642); \codestral{} (0.474) and \deepseekcoder{} (0.568) trail. The \texttt{bar\_count} number understates the failure only slightly: within $\pm 10$ bars accuracy is still 6--18\%, and median absolute error is 37--56 bars on a 1--1508 gold range, with models defaulting to typical phrase-length integers (16, 24, 32, 40) regardless of the score. On EMOPIA, every model reduces the four Russell quadrants to a near-binary classifier (\qwencoder{} predicts Q1 for 97/120 inputs; \deepseekcoder{} splits Q1/Q4), and none recovers the valence axis from raw LilyPond; \codestral{} produces no valid \texttt{[0--3]} index and collapses to 0, a format-adherence failure consistent with its \texttt{metadata\_prediction} drop (0.083 vs.\ 0.63--0.65 for the other models). Error detection without chain-of-thought is essentially out of reach: the best macro-F1 across all five corruption categories is 0.068 (\deepseekcoder{}, \texttt{invalid\_metadata}), and even nonsensical metadata (\texttt{\textbackslash key xx \textbackslash nonsense}) goes undetected, since the task requires triaging the corruption type and locating it in one shot, in an output format (space-separated bar indices) that none of the backbones adopt naturally.

\begin{table}[t]
\centering
\scriptsize
\setlength{\tabcolsep}{3pt}
\begin{tabular}{llcccc}
\toprule
Category & Task & \phifour{} & \qwencoder{} & \codestral{} & \deepseekcoder{} \\
\midrule
\multirow{2}{*}{Basic}
 & bar\_count            & 0.010 & 0.020 & 0.030 & 0.000 \\
 & metadata\_qa          & \textbf{0.767} & \textbf{0.767} & 0.683 & 0.667 \\
\midrule
\multirow{3}{*}{Segment}
 & bar\_sequencing       & \textbf{0.545} & 0.521 & 0.503 & 0.542 \\
 & next\_bar\_prediction & \textbf{0.521} & 0.420 & 0.370 & 0.454 \\
 & metadata\_prediction  & \textbf{0.650} & 0.633 & 0.083 & 0.517 \\
\midrule
\multirow{5}{*}{Sequence}
 & music\_captioning     & 0.903 & \textbf{0.935} & 0.839 & 0.677 \\
 & composer\_recognition & 0.875 & \textbf{0.885} & 0.500 & 0.760 \\
 & genre\_recognition    & 0.939 & \textbf{0.955} & 0.788 & 0.924 \\
 & emotion\_recognition  & 0.283 & 0.300 & 0.000 & \textbf{0.417} \\
 & error\_detection      & 0.009 & 0.005 & 0.014 & \textbf{0.041} \\
\midrule
\multicolumn{2}{l}{\textbf{Mutopia macro avg}} & \textbf{0.651} & 0.642 & 0.474 & 0.568 \\
\multicolumn{2}{l}{\textbf{Mutopia weighted avg}} & \textbf{0.626} & 0.611 & 0.461 & 0.566 \\
\bottomrule
\end{tabular}
\caption{Understanding results on the ten ABC-Eval-derived tasks. Metrics follow Table~\ref{tab:understanding-tasks}. Chance baselines: 0.25 for 4-way MC; effectively 0 for \texttt{bar\_count} and \texttt{error\_detection}. Bold marks the per-task best.}
\label{tab:understanding-results}
\end{table}

\section{Results}
\label{sec:results}
JS reads MusPy descriptors from rendered MIDI, FMD reads LilyBERT embeddings from raw LilyPond text. The two metrics agree when alignment to BMdataset is genuine: \phifour{} few-shot improves over zero-shot on both JS(test) and FMD(test) (0.933 to 0.701), and \codestral{} few-shot achieves the lowest in-domain FMD (0.696) with a higher JS than its zero-shot baseline. They diverge in two opposite directions. When a model collapses to a narrow but well-formed subdistribution, JS rewards it while FMD penalises it: \codestral{} with the hand-written A-minor demos scores the table's maximum JS(test) (89.44) yet its FMD is worse than zero-shot, because the MusPy descriptors land near corpus means in A minor while LilyBERT detects the surface form as foreign. The opposite happens for \deepseekcoder{} zero-shot, where FMD(test) is second-best in the row (0.887) but JS(test) is only 55.39: the model reproduces LilyPond surface form better than the rendered musical statistics. JS is interpretable but exploitable by narrow outputs; FMD is compile-independent and harder to fool, but does not localise which musical property changed. We therefore read both alongside compile rate, instead of fusing them into a single rank.

Tables~\ref{tab:matrix} and~\ref{tab:understanding-results} share the same four backbones, so the two halves compare directly. Generation strength does not track understanding strength. \phifour{} and \qwencoder{} lead on both, but \codestral{}, the best zero-shot compiler (79.3\%), is the weakest understander (0.474), with most of the gap on the format-sensitive tasks (\texttt{metadata\_prediction}, \texttt{emotion\_recognition}). The asymmetry matches the modality-bias diagnostics in Audio-LLMs~\cite{morais2025modality}: fluency can decouple from semantic competence.

Within understanding, the recognition--reasoning gap is sharper for LilyPond than for ABC. ABC-Eval reports about 90\% on basic syntax and about 25\% on sequence-level reasoning for the best proprietary system~\cite{zhao2026abceval}; on LilyPond, our open-weight 14--22~B models reach 0.68--0.96 on three recognition tasks (\texttt{music\_captioning}, \texttt{composer}, \texttt{genre}), but never lift \texttt{bar\_count} above 0.03 exact or \texttt{error\_detection} above 0.041 macro-F1. The harder structural tasks survive the move to a richer notation, consistent with prior work~\cite{yuan2024chatmusician,zhou2024canllmsreason}.

\section{Conclusion}
\lilybench{} is the first symbolic music benchmark built on LilyPond and the first unified evaluation of LLM music generation and understanding on a shared backbone set. The four open-weight models produce executable LilyPond zero-shot, recognise composer, genre, and captions at 0.68--0.96 accuracy, and adapt to demonstrations when given training-distribution exemplars. The methodological contribution is a direct comparison of Jensen--Shannon descriptor similarity and LilyBERT-based Fr\'echet Music Distance on the same reference corpus: the two metrics agree when alignment is genuine and diverge in two complementary directions, motivating triangulation with compile rate rather than fusion. We release the prompt bank, understanding benches, synthetic corruption set, and evaluation code for full reproducibility. Natural extensions include chain-of-thought and reasoning-focused models~\cite{zhou2024canllmsreason} on the structural tasks (especially \texttt{error\_detection}, where ABC-Eval~\cite{zhao2026abceval} reports about a 40\% gain), broadening the corpus beyond Baroque so that BMdataset and LilyBERT do not over-determine the FMD reference, and adding audio-LLM modality~\cite{morais2025modality} to test whether the recognition--reasoning gap narrows when listening is paired with reading.


\begin{acknowledgments}
We thank Mario Bolognani and the BaroqueMusic.it community for granting access to their LilyPond transcriptions.
\end{acknowledgments}

\section*{Declaration on Generative AI}
\label{declaration}
During the preparation of this work, the authors used the tool  Claude 4.7 in order to: improve writing style, grammar and spelling check. After using this tool, the authors reviewed and edited the content as needed and take full responsibility for the publication’s content. 

\bibliography{references}

\end{document}